\documentclass[3p,10pt,authoryear]{elsarticle}

\usepackage{amsmath,bm,epsfig}
\newcommand{\B}[1]{{\bm{#1}}}
\newcommand{\C}[1]{{\mathcal{#1}}}
\newcommand{\pa}{\partial}
\usepackage[latin1]{inputenc}
\usepackage{natbib}

\begin{document}

\begin{frontmatter}

\title{Viscoelastic Fracture of Biological Composites}

\author{Eran Bouchbinder$^1$ and Efim A. Brener$^{1,2}$}
\address{$^1$ Chemical Physics Department, Weizmann Institute of Science, Rehovot 76100, Israel\\
$^2$ Peter Gr\"unberg Institut, Forschungszentrum J\"ulich, J\"ulich 52425 Germany}
\date{\today}
\begin{abstract}
Soft constituent materials endow biological composites, such as bone, dentin and nacre, with viscoelastic properties that may play an important role in their remarkable fracture resistance. In this paper we calculate the scaling properties of the quasi-static energy release rate and the viscoelastic contribution to the fracture energy of various biological composites, using both perturbative and non-perturbative approaches. We consider coarse-grained descriptions of three types of anisotropic structures: (i) Liquid-crystal-like composites (ii) Stratified composites (iii) Staggered composites, for different crack orientations. In addition, we briefly discuss the implications of anisotropy for fracture criteria. Our analysis highlights the dominant lengthscales and scaling properties of viscoelastic fracture of biological composites. It may be useful for evaluating crack velocity toughening effects and structure-dissipation relations in these materials.
\end{abstract}


\end{frontmatter}

\section{Introduction}
\label{intro}

Biological composites, such as bone, dentin and nacre, exhibit outstanding mechanical properties \citep{Arzt2003, Peterlik2006a, Fratzl2007a, Ritchie2009, Dunlop2010, Ritchie2010, Launey2010, Ji2010}. In particular, they combine elastic stiffness and fracture resistance that is not yet achieved by synthetic composites of similar composition. Therefore, extensive recent efforts have been aimed at exploring the basic principles underlying the heterogeneous structures and deformation mechanisms of these materials, seeking guidelines for the development of novel synthetic composites \citep{Fratzl2007, Munch2008, Antonietti2010, Dunlop2010, Ji2010}. At the nano-scale, many biological composites consist of hard, plate-like, mineral crystals embedded in a soft protein matrix. The scale and spatial arrangement of the plate-like mineral crystals are believed to play a crucial role in endowing biological composites with their remarkable mechanical properties. For example, the nanometric dimensions of the mineral crystals in bone-like composites have been proposed to be fundamentally linked to the fracture resistance of these materials \citep{Arzt2003}. Furthermore, such biological nano-composites where shown to exhibit a generic nano-structure in which the hard mineral crystals are arranged in a parallel staggered pattern inside the soft protein matrix \citep{Fratzl2007, Dunlop2010, Launey2010, Ji2010}. Other nano-composites such as nacre, composed of parallel stratified arrays of hard mineral crystals, are of great interest \citep{Dunlop2010}. Our focus in this paper is on the macroscopic implications of these nano-structures and their constitutive behaviors.

Recent effort has been devoted to the experimental characterization and modeling of the dependence of the fracture resistance of biological composites on the crack length, the so-call ``crack extension resistance curve'' (R-curve) behavior \citep{Nalla2003, Nalla2004, Nalla2005, Kinney2005, Koester2008, Launey2009, Launey2010}. To the best of our knowledge, much less attention has been given to the increase in fracture resistance due to the finite velocity of cracks, in spite of the fact that even small velocities may generate a non-negligible contribution to the fracture resistance \citep{Sasaki1993, Okumura2002a, Okumura2003, Ji2004, Iyo2004}. In fact, the difference between the fracture toughness of hydrated and dehydrated biological composites may provide indirect evidence in favor of this possibility \citep{Kruzic2003}. This finite velocity effect can be attributed to the viscous component of the mechanical response of the soft constituent materials in biological composites \citep{Puxkandl2002, Hazenberg2006}. Our goal in this paper is to explore this possible toughening mechanism in the framework of a single timescale linear viscoelastic model for various nano-structures. We also consider fracture initiation and some related anisotropic effects.

To quantify fracture resistance, consider a crack moving at a velocity $v$ and write the total energy release rate $G_{tot}(v)$, i.e. the amount of energy needed to create a unit of crack surfaces, as
\begin{equation}
\label{Gtot}
G_{tot}(v) = G_0 + G_{vis}(v) \ ,
\end{equation}
where $G_0$ is the velocity-independent energy release rate and $G_{vis}(v)$ is the finite-$v$ dissipation associated with viscous deformation. Note that for $v\!\ge\!0$ the energy release rate equals the fracture energy, $\Gamma_{tot}(v)$, which is a {\em materials property}, and hence $G_0$ should be identified with the critical energy release rate $G_c$, i.e the fracture energy at the initiation of crack propagation. Nevertheless, we prefer to use the notation $G_0$ in Eq. (\ref{Gtot}) since it allows interpreting our results even when $G_0\!<\!G_c$, i.e. under sub-critical conditions, where $G_0$ is the elastic energy release rate associated with a virtual incremental extension of the crack. Since $G_{vis}(0)\!=\!0$, we can expand $G_{vis}(v)$ to the lowest order in $v$ as
\begin{equation}
\label{Gvis}
G_{vis}(v) \simeq  G_0 \frac{v\,\tau}{d_c} w(\ell/d_c) + {\C O}(v^2)\ ,
\end{equation}
where $\tau$ is a typical viscous relaxation timescale, $d_c$ is the smallest scale cutoff for a continuum description in a given problem, $\ell$ is a quantity of the dimension of length and $w(\cdot)$ is a dimensionless function. Note that even though biological composites may exhibit a hierarchy of viscous relaxation times, throughout this paper we adopt the simplifying assumption that there exists only one relevant viscous relaxation time, associated with the soft constituent materials in biological composites. Generalization to several viscous relaxation times is rather straightforward.

The generic two-dimensional fracture problem we consider in this paper consists of a crack of linear size $L$ within a large body (i.e. a body whose linear dimensions are much larger than $L$) and under an applied remote tensile stress $\sigma_\infty$ that tends to open it. Dimensional analysis implies that we can write $G_0$ as
\begin{equation}
\label{G0}
G_0 = \frac{\sigma_\infty^2\,L}{E} g(\bar\ell/L) \ ,
\end{equation}
where $E$ is a relevant elastic modulus, $g(\cdot)$ is a dimensionless function and $\bar\ell$ is a lengthscale. As implied above, the dimension of $G_0$ is energy per unit area.

The ultimate goal of this paper is to calculate $G_{vis}$, either perturbatively as in Eq. (\ref{Gvis}) or non-perturbatively as in Eq. (\ref{Gtot}), and $G_0$ in Eq. (\ref{G0}) for various composite structures. Our strategy in achieving this goal, which was strongly influenced by the work of de Gennes and Okumura \citep{Gennes1990, Gennes2000, Okumura2001, Okumura2002a, Okumura2002, Okumura2003, Okumura2005}, is to write down coarse-grained linear elastic energy functions and viscoelastic dissipation functions for biological composites of various structures, and to use available quasi-static crack solutions to estimate in a perturbative manner the small-velocity linear viscoelastic contribution to the fracture energy. We then use ``matched asymptotics'' considerations to show how this analysis can be extended to a wider range of crack velocities. We focus on the scaling properties of these quantities, i.e. we systematically neglect pre-factors of order unity, and consider three types of anisotropic structures: (i) Liquid-crystal-like composites (ii) Stratified composites (iii) Staggered composites, for different crack orientations with respect to the internal structure.

In ordinary isotropic viscoelastic fracture (the details are provided in section \ref{isotropic}) $\ell$ scales with a microscopic cutoff length (usually termed the ``process zone size'' \citep{Broberg1999}), $E$ scales with the isotropic elastic modulus and $\bar\ell$ scales with a macroscopic (geometric) cutoff length, the crack's length $L$ for the fracture configuration considered here, and the functions $g(\cdot)$ and $w(\cdot)$ are of order unity. This is the hallmark of isotropic fracture: large scales elastic energy is dissipated at the small scales near the tip of a crack. Our results, summarized in Tables 1-3, show that the presence of anisotropic nano-structures introduces additional lengthscales (determined, for example, by the ratio of the elastic moduli of the soft and hard constituent materials or by the aspect ratio of the plate-like mineral crystals) gives rise to different scaling behavior as compared to isotropic fracture. In addition, we show that anisotropy may have some implications for fracture criteria.

The structure of this paper is as follows. In Section \ref{procedure_isotropic} we describe the general procedure we adopt and apply it to isotropic viscoelastic fracture. In Section \ref{LC} we consider fracture in liquid-crystal-like structures. In Section \ref{stratified} we consider fracture in stratified (layered) composite, for two crack orientations (parallel and perpendicular to the  layers), while in Section \ref{staggered} we consider fracture in staggered composites. Section \ref{fracture_criteria} briefly discusses anisotropy effects and their relevance for fracture criteria. In Section \ref{velocities} we show how to extend the perturbative approach to a wider range of finite crack velocities. Section \ref{summary} offers a summary and some concluding remarks.

\section{General procedure and application to isotropic viscoelastic fracture}
\label{procedure_isotropic}

\subsection{General procedure}
\label{procedure}

In order to calculate $G_0$ and $G_{vis}$ for various nano-structures we describe first the general procedure we follow. In this Section we focus on a perturbative approach, while in Section \ref{velocities} we show how to extend the analysis to the non-perturbative, finite $v$, regime. As stressed above, we have enormously benefitted from the papers of de Gennes and Okumura \citep{Gennes1990, Gennes2000, Okumura2001, Okumura2002a, Okumura2002, Okumura2003, Okumura2005} and in many ways the present contribution is a further development of their work. The starting point of our procedure is a nano-mechanical model that incorporates the salient features of a given nano-structure into a continuum, coarse-grained, description of the effective viscoelastic response of a composite material. Specifically, this crucial step results in a coarse-grained elastic energy density function $f(\nabla \B u, \{E_i\})$ and a coarse-grained dissipation function $R(\nabla \dot{\B u}, \{\eta_i\})$, where $\B u$ is the coarse-grained displacement field and $\{E_i, \eta_i\}$ is a set of elastic moduli and viscosity coefficients (respectively) associated with the different constituent materials in a given nano-composite. Note that the dimension of $f$ is energy per unit volume and of $R$ is energy per unit length per unit time.

The next step in the perturbative approach is to derive an equation of motion for $\B u$ by looking for the stationary variation of $f(\nabla \B u, \{E_i\})$ with respect to $\B u$ and neglecting $R(\nabla \dot{\B u}, \{\eta_i\})$. Then one should solve the equation for $\B u$ in the presence of a crack, i.e. a line that cannot support tensile and shear traction, under the external loading conditions. In many cases scaling arguments and matched asymptotics considerations can be useful in obtaining the scaling properties of $\B u$ in different regions in space.

\begin{figure}
\centering \epsfig{width=.4\textwidth ,file=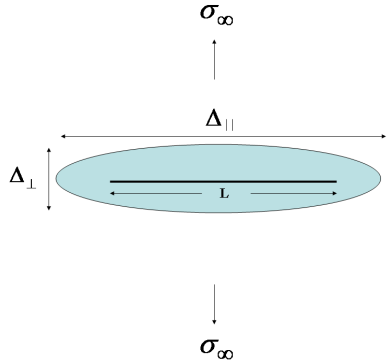}
\caption{A schematic sketch of a crack of length $L$ in an infinite medium (i.e. a medium of linear dimensions much larger than $L$) remotely loaded by a uniform tensile stress $\sigma_\infty$.
The shaded area represents the typical strain distribution around the crack, where $\Delta_\bot$ is the scale in the direction perpendicular to the crack and $\Delta_\|$ is the scale in the direction parallel to the crack. The relative magnitude of $\Delta_\bot$ and $\Delta_\|$ will vary from problem to problem and is not necessarily as shown schematically in the figure.}\label{sketch}
\end{figure}

The final step is to calculate $G_0$ and $G_{vis}$ as follows. The two-dimensional crack configuration we described above is sketched in Fig. \ref{sketch}. The crack is assumed to propagate at a velocity $v$ which is much smaller than the speed of sound. Recall that the presence of a crack is expressed by the usual mixed boundary conditions
\begin{eqnarray}
\label{BC}
\pa_\bot u_\bot &=& 0 \quad\hbox{on the crack}  \ ,\\
u_\bot &=& 0 \quad\hbox{along the symmetry line, outside the crack} \ .
\end{eqnarray}

The shaded area represents the typical strain distribution around the crack, where $\Delta_\bot$ is the scale in the direction perpendicular to the crack and $\Delta_\|$ is the scale in the direction parallel to the crack. The relative magnitude of $\Delta_\bot$ and $\Delta_\|$ will vary from problem to problem and in not necessarily as is shown schematically in the figure.
The energy release rate $G_0$ is just the elastic strain energy density ahead of the crack tip multiplied by the spatial extent of the strain distribution in the direction perpendicular to crack propagation. Therefore, following Fig. \ref{sketch}, we can immediately write it as
\begin{equation}
\label{G_0}
G_0 \sim \frac{\sigma^2_\infty}{E_\bot} \Delta_\bot \ ,
\end{equation}
where $E_\bot$ is the elastic modulus in the perpendicular direction. Comparing Eq. (\ref{G_0}) with Eq. (\ref{G0}) implies that $E$ in the latter is $E_\bot$ and that $\Delta_\bot\!\sim\!L g(\bar\ell/L)$.
The strain in the perpendicular direction, on a scale $\Delta_\bot$, is easily estimated as
\begin{equation}
\label{strain}
\frac{u_\bot}{\Delta_\bot} \sim \frac{\sigma_\infty}{E_\bot} \ .
\end{equation}

%%%%%%% FIGURE 1 %%%%%%%%%%%%%%%%%%
%%%%%%%%%%%%%%%%%%%%%%%%%%%%%%%%%%%

Finally, the fracture viscous dissipation $G_{vis}$ is given as
\begin{equation}
\label{G_vis}
G_{vis} = \frac{R(\nabla \dot{\B u}, \{\eta_i\})}{v} \ .
\end{equation}
Under steady state propagation conditions we can replace time derivative in $R(\nabla \dot{\B u}, \{\eta_i\})$ by space derivatives following $\pa_t = -v \pa_\|$, where $\|$ denotes the direction of crack propagation. The procedure described here allows one to systematically calculate $G_0$ and $G_{vis}$ in a large class of problems. It is important to note that this procedure is perturbative in nature since we calculate the static crack solution disregarding viscous deformation, and use it later to estimate the dissipation arising from viscous deformation. A non-perturbative extension of this procedure is discussed in Section \ref{velocities}.

\subsection{Isotropic linear viscoelastic fracture}
\label{isotropic}

In order to demonstrate how the procedure described above actually works and also to set a reference case for the anisotropic problems to be considered later, we start by discussing ordinary isotropic linear viscoelastic fracture. Isotropy implies that
\begin{equation}
E_\bot, E_\| \sim E, \quad u_\bot, u_\| \sim u, \quad \Delta_\bot, \Delta_\| \sim L \ ,
\end{equation}
where $u$ is the displacement in either the $x$ or $y$ directions and $E$ is the isotropic elastic modulus.
According to the first step in our procedure, we need to write down an expression for $f$ and $R$. The elastic energy density has the form (scaling-wise)
\begin{equation}
f \sim E (\pa u)^2 \ ,
\end{equation}
where the spatial derivative $\pa$ refers to either $x$ or $y$.
In addition, the viscous dissipation function reads
\begin{equation}
R \sim \eta \int (\pa \dot u)^2 r dr d\theta\ ,
\end{equation}
where $\eta$ is the viscosity, $(r, \theta)$ is a polar coordinates system whose origin is at the tip of the crack and the two-dimensional integration extends over the whole body.

The next step is to derive an equation of motion for $\B u$ by looking for the stationary variation of $f$ with respect to $\B u$ and to solve it in the presence of a crack. It is well known that the resulting equation is the Lam\'e equation and that the solution at intermediate scales reads \citep{Broberg1999}
\begin{equation}
\label{LEFM}
u \sim \frac{\sigma_\infty \sqrt{L~r}}{E} \ ,
\end{equation}
which implies the well-known square-root singularity for the stress, $\sigma\!\sim\! \sigma_\infty \sqrt{L/r}$.
Intermediate scales mean scales between a microscopic inner cutoff lengthscale, say $d_c$, where the linear theory fails (the so-called ``process zone scale'') and the outer scale $L$. As a consistency check we note that at $r\!\sim\!L$, we recover Eq. (\ref{strain}) (recall that $\Delta_\bot\!\sim\!L$) and $\sigma\!\sim\!\sigma_\infty$ as expected.

The last step is to use Eqs. (\ref{G_0}) and (\ref{G_vis}) to estimate $G_0$ and $G_{vis}$. Substituting $\Delta_\bot\!\sim\!L$ in Eq. (\ref{G_0}), we immediately obtain
\begin{equation}
\label{G_isotropic}
G_0 \sim \frac{\sigma^2_\infty L}{E} \ ,
\end{equation}
We now substitute Eq. (\ref{LEFM}) in Eq. (\ref{G_vis}) to obtain
\begin{eqnarray}
\label{dissipation LEFM}
G_{vis} \sim \frac{\eta}{v} \int_{d_c}^L (\pa_r \dot{u})^2 r dr \sim E ~v~ \tau \int_{d_c}^L (\pa_{rr} u)^2 r dr \sim  \frac{\sigma^2_\infty L}{E} ~v~ \tau \int_{d_c}^L \frac{r dr}{r^3}\sim G_0  ~v~ \tau \left(\frac{1}{{d_c}}-\frac{1}{L} \right) \sim G_0 \frac{v~\tau}{{d_c}} \ ,
\end{eqnarray}
where $d_c \ll L$ and $\tau\! \equiv \!\eta/E$.

Comparing Eqs. (\ref{G_isotropic}) and (\ref{dissipation LEFM}) with Eqs. (\ref{Gvis}) and (\ref{G0}) we immediately observe that $w\!\sim\!{\C O}(1)$ and $g\!\sim\!{\C O}(1)$, i.e. in isotropic fracture the viscous dissipation is controlled by a microscopic lengthscale $\ell \!\sim\! d_c$ and quasi-static energy release rate is controlled by a macroscopic lengthscale $\bar\ell\!\sim\!L$. This is the hallmark of isotropic fracture: energy is released from large scales and is dissipated at the small scales near the tip of a crack \citep{Broberg1999}.

\section{Liquid-Crystal-like structures}
\label{LC}

The structure of smectic liquid crystals is somewhat reminiscent of the structure of some biological composites, as was previously noted in \citet{Okumura2002}. We therefore start our discussion of anisotropic composite structures by considering liquid-crystal-like structures. By that we mean a structure that is composed of ordered layers of width $d$ that lie along, say, the $x$-direction and can deform along the $y$-direction, by both stretching and bending. No elastic deformation along the $x$-direction takes place. Furthermore, gradients of the displacement rate (material velocity) give rise to a viscous response. Such a material is characterized by the following elastic energy functional \citep{Gennes1990}
\begin{equation}
\label{elastic_smectic}
f \sim E (\pa_y u_y)^2 + E d^2 (\pa_{xx} u_y)^2
\end{equation}
Here $u_y(x,t)$ is the displacement, $E$ quantifies the elastic stiffness in the $y$-direction and $E d^2$ is associated with the bending stiffness of the layers. The viscous dissipation function of such a material takes the form
\begin{equation}
\label{dissipation_smectic}
R \sim \eta \int \left[(\pa_x \dot{u}_y)^2+(\pa_y \dot{u}_y)^2\right] dx dy \ ,
\end{equation}
where $\eta$ is the viscous response coefficient. This material response can approximately describe systems such as lamellar phases of block copolymers \citep{Kato2002}, but strictly not bone-like materials. Nevertheless, we believe that this example is very instructive and relevant in the present context.

The stationary variation of $f$ in Eq. (\ref{elastic_smectic}) with respect to $u_y$ reads \citep{Gennes1990}
\begin{equation}
\label{Eq_lc}
\pa_{yy} u_y - d^2 \pa_{xxxx} u_y = 0 \ ,
\end{equation}
which is of course valid on scales larger than $d$. The crack is parallel to the layers, i.e. located along the $x$-direction.
The competition between stretching and bending (i.e. the different order of the spatial derivatives) in Eq. (\ref{Eq_lc}) immediately implies anisotropic scaling. Specifically, we observe that
\begin{equation}
\Delta^2_x \sim d \Delta_y \ ,
\end{equation}
where $\Delta_x$ is $\Delta_\|$ of Fig. \ref{sketch} and $\Delta_y$ is $\Delta_\bot$ of Fig. \ref{sketch}.
Since we expect $\Delta_x\!\sim\!L \gg d$, we immediately deduce that
\begin{equation}
\label{LC_scaling}
\Delta_y \sim \frac{L^2}{d} \gg \Delta_x  \ .
\end{equation}
Substituting this result in Eq. (\ref{G0}) and identifying $E_\bot\!=\!E$ yields
\begin{equation}
G_0 \sim  \frac{\sigma^2_\infty L^2}{E d} \ ,
\end{equation}
which is identical to the result previously derived in \citet{Gennes1990}. We first observe that $G_0$ for liquid-crystal-like structures contains the microscopic scale $d$. Furthermore, it is a factor $L/d\!\gg\!1$ larger than the isotropic result of Eq. (\ref{G_isotropic}).

In order to calculate $G_{vis}$ we need the solution of Eq. (\ref{Eq_lc}) in the presence of a crack. This problem was considered in \citet{Gennes1990}, where it was found that
\begin{equation}
u_y \sim x \,h\!\left(\frac{x}{\sqrt{d\,y}} \right) \ .
\end{equation}
Focusing on the outer scale of the problem, i.e. $x\!\sim\!\Delta_x\!\sim\!L$ and $y\!\sim\!\Delta_y\!\sim\!L^2/d$, Eq. (\ref{strain}) immediately tells us that in fact
\begin{eqnarray}
u_y \sim \frac{\sigma_\infty L}{E d} x \,h\!\left(\frac{x}{\sqrt{d\,y}}\right) \ .
\end{eqnarray}

To estimate $G_{vis}$, we first note that
\begin{equation}
\label{dissipation_smecticA}
R \sim \eta v^2 \int \left[(\pa_{xx} u_y)^2+(\pa_{xy} u_y)^2\right] dx dy \sim \eta v^2 \int (\pa_{xx} u_y)^2 dx dy \ ,
\end{equation}
because $\Delta_y\!\gg\!\Delta_x$ (i.e. gradients in the x direction are much larger than in the y direction). Then, we obtain
\begin{equation}
\label{integral_lc}
G_{vis} = \frac{R}{v} \sim \eta v \int (\pa_{xx} u_y)^2 dx dy \sim  v\tau \frac{\sigma^2_\infty L^2}{E d^2} \int^{x \sim L} \int^{y \sim L^2/d} \frac{k\!\left(\frac{x}{\sqrt{d~ y}}\right)}{d~ y} dx dy \ ,
\end{equation}
where $k(\cdot)$ is related to $h(\cdot)$ and its derivatives, and $\tau\!=\!\eta/E$.
The function $k(\cdot)$ in the integrand is finite in the limit $y \to 0$. Therefore, the integrand is characterized by an integrable singularity that scales as $y^{-1/2}$ (easily seen by introducing an auxiliary variable $\tilde x \equiv x /\sqrt{d y}$) and the integral is dominated by the upper limits of integration, which satisfy the scaling relations discussed above. Finally, performing the integration we obtain
\begin{equation}
\label{dissipation smectic}
G_{vis} \sim  G_0 \frac{v \tau}{d} \frac{L}{d} \ .
\end{equation}
Comparing this result to Eq. (\ref{Gvis}) we observe that $d_c\!\sim\!d$, $w\!\sim\!L/d$ and $\ell\!\sim\!L$. This suggests that $G_{vis}$ is affected {\em by the large scale $L$}. It is remarkable that crack dissipation is controlled by the outer scale of a fracture problem. We believe that this unusual result derives from the infinite anisotropy of the liquid-crystal-like structures; those structures posses no elastic stiffness at all in the direction parallel to the layers. Finally, we note the if we interpret $d_c$ in Eq. (\ref{dissipation LEFM}) as the width the layers $d$, which is rather artificial, we can say that $G_{vis}$ for liquid-crystal-like structures is a factor $L/d \gg 1$ larger than the result for isotropic linear viscoelastic fracture.

\section{Stratified composites}
\label{stratified}

\begin{figure}
\centering \epsfig{width=.4\textwidth ,file=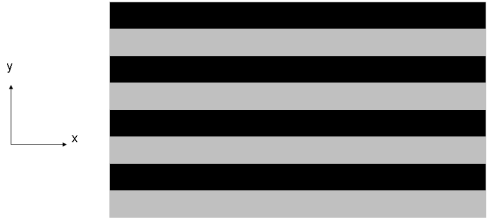}
\caption{A schematic sketch of a stratified structure composed of repeated parallel layers of width $d$. The darker layers are elastically hard, with isotropic stiffness $E_h$ and width $d$, and the brighter layers are elastically soft, with isotropic stiffness $E_s\!\ll\!E_h$ and width $d$.}\label{stratified_fig}
\end{figure}

Here we consider stratified (layered) biological nano-composites, see Fig. \ref{stratified_fig}. These structures are periodic in the $y$-direction and are translationally invariant in the $x$-direction. They are composed of hard layers of width $d$ and elastic modulus $E_h$ and soft layers of width $d$ and elastic modulus $E_s\!\ll\!E_h$. The effective Hooke's law for this structure can be calculated exactly, as presented in detail in Appendix \ref{appendix}. For a mineral volume fraction of the order of $1/2$ the result reads
\begin{equation}
\label{elastic_layered}
f \sim E_h (\pa_x u_x)^2 + E_s (\pa_y u_y)^2 + E_s (\pa_x u_x)(\pa_y u_y) + E_s (\pa_x u_y+\pa_y u_x)^2 + E_h d^2 (\pa_{xx} u_y)^2 \ ,
\end{equation}
where a bending term, identical to the one that appeared for the liquid-crystal-like structures, was included.  This expression is identical to the one proposed in \citet{Okumura2002}.

%%%%%%% FIGURE 2 %%%%%%%%%%%%%%%%%%
%%%%%%%%%%%%%%%%%%%%%%%%%%%%%%%%%%%

In order to obtain the viscous dissipation function, we assume that the hard material is purely elastic and that the soft material is linear viscoelastic with a viscous response coefficient $\eta_s$. Therefore, we obtain
\begin{equation}
\label{dissipation_layered}
R \sim \eta_s \int \left[(\pa_y \dot{u}_y)^2 + (\pa_x \dot{u}_x)(\pa_y \dot{u}_y) + (\pa_x \dot{u}_y+\pa_y \dot{u}_x)^2\right] dx dy \ .
\end{equation}
Unlike the liquid-crystal-like structures, stratified composite structures have elastic stiffness in both the $x$ and the $y$ directions. Therefore, we should (at least) distinguish between cracks that are parallel and perpendicular to the layers.

\subsection{Parallel cracks}

We consider a crack located along the $x$-direction, i.e. parallel to the layers. The presence of a new small parameter
\begin{equation}
\label{eps}
\epsilon \equiv \frac{E_s}{E_h} \ll 1
\end{equation}
implies a new lengthscale in the problem
\begin{equation}
\lambda \equiv \frac{d}{\sqrt\epsilon} \gg d \ .
\end{equation}
We focus on the limit
\begin{equation}
d \ll \lambda \ll L \ .
\end{equation}
As was shown in \citet{Okumura2002}, in this case the elastic functional of Eq. (\ref{elastic_layered}) cannot be approximated by a single expression for all relevant scales. We therefore consider separately large scales, $x\!\gg\!\lambda$, and small scales, $x\!\ll\!\lambda$. Consider first the large scales. Following \citet{Okumura2002}, the elastic functional of Eq. (\ref{elastic_layered}) can be approximated by
\begin{equation}
\label{f_large_stratified}
f \sim E_s (\pa_x u_y)^2 + E_s (\pa_y u_y)^2 \ .
\end{equation}
This implies isotropic, mode III-like, fracture in a soft material. We therefore use the results of isotropic fracture, now with a lower cutoff scale $\lambda$
\begin{eqnarray}
G_0 &\sim& \frac{\sigma^2_\infty L}{E_s}, \\
G^{(l)}_{vis} &\sim& G_0 \frac{v \tau}{\lambda} \ ,
\end{eqnarray}
where the superscript $(l)$ denotes large scales.

For the small scales, the elastic functional of Eq. (\ref{elastic_layered}) can be approximated by
\begin{equation}
\label{stratified_small}
f \sim E_s (\pa_y u_y)^2 + E_h d^2 (\pa_{xx} u_y)^2 \ .
\end{equation}
To proceed, we employ ``matched asymptotics'' considerations, i.e. we determine the properties of the ``inner'' (small scales) solution by demanding that it smoothly crosses over to the ``outer'' (larger scales) solution at a scale $x\!\simeq\!\lambda$. Therefore, the inner problem is characterized by a loading $\sigma^{(\!s)}_\infty\!\sim\!\sigma_\infty \sqrt{L/\lambda}$, with
\begin{eqnarray}
E^{(\!s)}_\bot = E_s, \quad (\Delta^{(\!s)}_x)^2 \simeq \lambda \Delta^{(\!s)}_y, \quad \Delta^{(\!s)}_x \simeq \Delta^{(\!s)}_y \simeq \lambda \ .
\end{eqnarray}
Here the superscript $(s)$ denotes small scales, i.e. all of the quantities refer to the region near the tip where the energy functional in Eq. (\ref{stratified_small}) dominates the elastic behavior and not to the shaded region in the global problem sketched in Fig. \ref{sketch}.

$G_0$ for the small scales must be identical to that of the large scales as no dissipation takes place in the intermediate region. Using Eq. (\ref{strain}), we obtain
\begin{eqnarray}
u_y \sim \frac{\sigma_\infty}{E_s} \sqrt{\frac{L}{\lambda}} \, x \,h\!\left(\frac{x}{\sqrt{\lambda \,y}}\right) \ ,
\end{eqnarray}
where we used $x\!\sim\!\Delta_x\!\sim\!\lambda$. We then note that
\begin{equation}
R \sim \eta_s \int \left[(\pa_y \dot{u}_y)^2 + (\pa_x \dot{u}_x)(\pa_y \dot{u}_y) + (\pa_x \dot{u}_y+\pa_y \dot{u}_x)^2\right] dx dy \simeq E_s \tau v^2 \int (\pa_{xy} u_y)^2 dx dy \ ,
\end{equation}
because $\Delta_y\!\ll\!\Delta_x$ (i.e. gradients in the y direction are much larger than in the x direction).
For $G^{(\!s)}_{vis}$ we obtain
\begin{equation}
G^{(s)}_{vis} \sim E_s \tau v \int (\pa_{xy} u_y)^2 dx dy \sim E_s \tau v \frac{\sigma^2_\infty L}{E_s^2 \lambda} \int_{x \sim \sqrt{\lambda\,d}} \int_{y \sim d} \frac{k\!\left(\frac{x}{\sqrt{d\,y}}\right)}{y^2}
dx dy \ ,
\end{equation}
where $k(\cdot)$ is related to $h(\cdot)$ and its derivatives.
The function $k(\cdot)$ in the integrand is finite in the limit $y \to 0$. Therefore, the integrand is characterized by a non-integrable singularity that scales as $y^{-3/2}$ (easily seen by introducing an auxiliary variable $\tilde x \equiv x /\sqrt{d y}$) and the integral is dominated by the lower limits of integration, which satisfy the scaling relations discussed above. Finally, performing the integration we obtain
\begin{equation}
G^{(s)}_{vis} \sim G_0 \frac{v \tau}{\sqrt{d\,\lambda}} \ .
\end{equation}
The overall $G_{vis}$ is then given as
\begin{equation}
\label{dissipation startified1}
G_{vis}=G^{(l)}_{vis}+G^{(s)}_{vis} \sim G_0 \frac{v \tau}{\lambda}\left( 1+ \sqrt{\frac{\lambda}{d}}\right) \simeq G_0 \frac{v \tau}{\sqrt{d\,\lambda}} \ .
\end{equation}
%Note that since $G_{vis}$ is dominated by the small scales dissipation in this case, it is identical (scaling-wise) to $G_{vis}$ obtained in Eq. (\ref{dissipation startified}) for the case $d\!\ll\!L\!\ll\!\lambda$.

It is interesting to note that the result in Eq. (\ref{dissipation startified1}) can be written as
\begin{equation}
\label{d_parallel}
G_{vis} \sim  G_0 \frac{v \tau}{d_\|} \ ,
\end{equation}
where $d_\|$ is the smallest cutoff length in the crack propagation direction. In the present case the crack propagates in the $x$-direction and we have $d_\|\!=\!d_x\!\sim\!\sqrt{d\,\lambda}$. As we will see below, this result is rather generic as long as the dissipation integral is dominated by the lower limits of integration.

\subsection{Perpendicular cracks}

Here we consider a crack located along the $y$-direction, i.e. perpendicular to the layers. Following \citet{Okumura2001}, in this case the elastic functional of Eq. (\ref{elastic_layered}) can be approximated as
\begin{equation}
f \sim E_h (\pa_x u_x)^2 + E_s (\pa_y u_x)^2 \ .
\end{equation}
Minimizing the elastic energy with respect to $u_x$, we obtain the following equation
\begin{equation}
\label{eom_perp}
\pa_{xx} u_x + \pa_{\tilde y \tilde y} u_x = 0 \ ,
\end{equation}
where $\tilde y \!\equiv\! \epsilon^{-1/2}y$ (recall that $\epsilon\!=\!E_s/E_h\!\ll\!1$). This is analogous to a mode-III crack problem \citep{Broberg1999}.

Note that for the crack orientation considered here $\Delta_\bot\!=\!\Delta_x$, $\Delta_\|\!=\!\Delta_y$ and $E_\bot\!=\!E_h$. Since we expect $\Delta_y \!\simeq\! L$, we immediately obtain
\begin{eqnarray}
\Delta_x \simeq \epsilon^{-1/2} L \gg \Delta_y \ .
\end{eqnarray}
Substituting in Eq. (\ref{G_0}), we obtain
\begin{equation}
\label{G_perp}
G_0 \sim \frac{\sigma^2_\infty L}{E_h \epsilon^{1/2}}  = \frac{\sigma^2_\infty L}{\sqrt{E_h E_s}}\ .
\end{equation}
This result shows that the energy release rate in this case is significantly {\em enhanced} as compared to fracture in an isotropic material with stiffness $E_h$.

We consider now Eq. (\ref{strain}) which tells us that the crack opening on a scale $y\!\sim\!L$ reads
\begin{equation}
u_x \sim \frac{\sigma_\infty}{E_h} \epsilon^{-1/2} L \ .
\end{equation}
Furthermore, from Eq. (\ref{eom_perp}) we know that the opening displacement should satisfy ordinary fracture scaling in the coordinates $(x,\tilde y)$, i.e.
\begin{equation}
u_x \sim (x^2 + \tilde y ^2)^{1/4} = (x^2 + \epsilon^{-1} y ^2)^{1/4} \ .
\end{equation}
Therefore, we conclude that
\begin{equation}
\label{ux_perp}
u_x \sim \frac{\sigma_\infty}{E_h} \epsilon^{-1/4} \sqrt{L} (x^2 + \epsilon^{-1} y ^2)^{1/4} \ .
\end{equation}
We can now use the above expression for $u_x$ to calculate the viscous dissipation. We first note that
\begin{equation}
\label{R_perp_layered}
R \sim \eta_s \int \left[(\pa_y \dot{u}_y)^2 + (\pa_x \dot{u}_x)(\pa_y \dot{u}_y) + (\pa_x \dot{u}_y+\pa_y \dot{u}_x)^2\right] dx dy \simeq E_s \tau v^2 \int (\pa_{yy} u_x)^2 dx dy \ ,
\end{equation}
because $\Delta_y\!\ll\!\Delta_x$ (i.e. gradients in the y direction are much larger than in the x direction) and $\pa_t\!=\!-v \pa_y$. Therefore, we obtain
\begin{equation}
\label{diss_perp_layered1}
G_{vis} = \frac{R}{v} \sim E_s \tau v \int (\pa_{yy} u_x)^2 dx dy \sim E_s \tau v \frac{\sigma^2_\infty L}{E^2_h \epsilon^{1/2}} \int_{x \sim \epsilon^{-1/2} d} \int_{y \sim d} \frac{k\!\left(\frac{x}{\epsilon^{-1/2}~ y}\right)}{\epsilon^{1/2} y^3} dx dy \ ,
\end{equation}
where we used the lower limits of integration alone because of the small scales divergence. The integral alone yields $(\epsilon\,d)^{-1}$, which implies that
\begin{equation}
\label{diss_perp_layered}
G_{vis} \sim E_s \tau v \frac{\sigma^2_\infty L}{E^2_h \epsilon^{1/2}} \frac{1}{\epsilon d} \sim G_0 \frac{v \tau}{d}  \ .
\end{equation}
Note that this result agrees with Eq. (\ref{d_parallel}), where this time we have $d_\|\!=\!d_y\!\sim\!d$.

\section{Staggered composites}
\label{staggered}

\begin{figure}
\centering \epsfig{width=.4\textwidth ,file=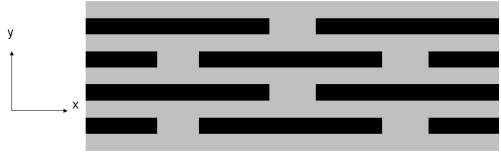}
\caption{A schematic sketch of a staggered array of hard platelets (darker) embedded within a soft matrix (brighter). The width of the platelets is $d$ and their length is $\rho d$, where $\rho$ is a dimensionless aspect ratio. The length of the horizontal (along $x$) gap between platelets in the same layer, filled with a soft material, is $a$. The width of the infinitely long soft layers is also $d$.}\label{staggered_fig}
\end{figure}

Micro-structural studies \citep{Fratzl2007, Dunlop2010} have demonstrated that bone-like materials feature a staggered arrangement of hard mineral platelets at the nano-scale, see Fig. \ref{staggered_fig}. The width of the platelets is $d$ and their length is $\rho d$, where $\rho$ is a dimensionless aspect ratio. The length of the horizontal (along $x$) gap between platelets in the same layer, filled with a soft material, is $a$. The width of the infinitely long soft layers is also $d$. Our goal here is to find out whether the staggered structure gives rise to a different viscous fracture dissipation as compared to stratified structures. The structural difference between the staggered array and the stratified one is that in the former the hard mineral platelets are of finite length, while in the latter they are infinite. This implies that the resulting stiffness in $x$-direction may be different. Therefore, denoting the elastic stiffness in this direction by $E_x^{eff}$, we can write
\begin{equation}
\label{elastic_staggered}
f \sim E_x^{eff} (\pa_x u_x)^2 + E_s (\pa_y u_y)^2 + E_s (\pa_x u_x)(\pa_y u_y) + E_s (\pa_x u_y+\pa_y u_x)^2 + E_h d^2 (\pa_{xx} u_y)^2 \ .
\end{equation}

%%%%%%% FIGURE 3 %%%%%%%%%%%%%%%%%%
%%%%%%%%%%%%%%%%%%%%%%%%%%%%%%%%%%%

In order to estimate $E_x^{eff}$, we adopt the Tension-Shear-Chain (TSC) nano-mechanical model \citep{Arzt2003, Ji2004, Gao2006}. The major assumption of this model is that tensile stresses are transferred between the hard mineral crystals mainly by shear stresses in the soft matrix. In other words, the tensile stress in the gap of length $a$ is assumed to be negligible. This assumption is valid in the limit of large stiffness contrast $\epsilon\!\ll\!1$, large aspect ratio of the hard platelets, $\rho\!\gg\!1$, and $a\!\ll\!\rho d$. In this limit, the model predicts the following expression for $E_x^{eff}$ \citep{Arzt2003}
\begin{equation}
\label{Eeff}
\frac{1}{E_x^{eff}} \simeq \frac{8(1+\nu_s)(1-\phi)}{E_s \phi^2 \rho^2} + \frac{1}{\phi E_h} \ ,
\end{equation}
where $\nu_s$ is the soft material Poisson's ratio and $\phi$ is the volume fraction of the hard material. We focus on $\phi\!\simeq\!1/2$. The first term on the right hand side of Eq. (\ref{Eeff}) represents the shear deformation in the soft material between the hard platelets, while the second one the tensile deformation inside the platelets.

We are mainly interested in the regime where $E_s\!\ll\!E_s \rho^2\!\leq \! E_h$, i.e. when the shear deformation of the soft material is important. Under this condition we can write
\begin{equation}
\label{Eeff1}
E_x^{eff} \simeq E_s  \rho^2 \ .
\end{equation}
The corresponding dissipation function reads
\begin{equation}
\label{dissipation_staggered}
R \sim \eta_s \int \left[\rho^2 (\pa_x \dot{u}_x)^2 + (\pa_y \dot{u}_y)^2 + (\pa_x \dot{u}_x)(\pa_y \dot{u}_y) + (\pa_x \dot{u}_y+\pa_y \dot{u}_x)^2\right] dx dy \ .
\end{equation}
Comparing this expression to Eq. (\ref{dissipation_layered}) for stratified materials, we immediately identify the additional term proportional to $\rho^2 (\pa_x \dot{u}_x)^2$. The physics behind this new contribution is clear; it represents the additional viscous dissipation due to shear deformation of the soft matrix during tensile deformation in the $x$-direction. Since $\rho\!\gg\!1$ is a large number, one may expect a qualitative change in the fracture dissipation.

In order to check if this is indeed the case, we first note that since for cracks parallel to the platelets (i.e. along the $x$-direction) the deformation is dominated by $u_y$, the results are identical to those for stratified structures. For cracks perpendicular to the platelets (i.e. in the $y$-direction), we compare Eqs. (\ref{elastic_staggered}) and (\ref{Eeff1}) with Eq. (\ref{elastic_layered}), which immediately suggests that we should simply identify $E_h$ is the latter with $E_s \rho^2$, and use all of the results obtained before for perpendicular cracks in stratified structures. In particular, we identify $\epsilon$ of Eq. (\ref{eps}) with $\rho^{-2}$ and obtain
\begin{eqnarray}
E_\bot\!=\!E_s \rho^2,\quad\Delta_x \!\simeq\! \rho \Delta_y,\quad\Delta_y \!\simeq\! L,\quad\Delta_x \!\simeq\! \rho L \gg \Delta_y \ ,
\end{eqnarray}
together with
\begin{equation}
\label{G_perp_stag}
G_0 \sim \frac{\sigma^2_\infty L}{E_s \rho}, \quad u_x \sim \frac{\sigma_\infty}{E_s \rho^{3/2}} \sqrt{L} (x^2 + \rho^2 y ^2)^{1/4} \ .
\end{equation}

We can now use the above expression for $u_x$ to calculate the viscous dissipation. We use Eq. (\ref{dissipation_staggered}) to obtain
\begin{eqnarray}
R &\sim&  \eta_s \int \left[\rho^2 (\pa_x \dot{u}_x)^2 + (\pa_y \dot{u}_y)^2 + (\pa_x \dot{u}_x)(\pa_y \dot{u}_y) + (\pa_x \dot{u}_y+\pa_y \dot{u}_x)^2\right] dx dy \nonumber\\
 &\sim& E_s \tau v^2 \int \left[\rho^2 (\pa_{xy} u_x)^2 + (\pa_{yy} u_x)^2 \right] dx dy \ ,
\end{eqnarray}
The second term in the above expression must result in the same contribution as in Eqs. (\ref{diss_perp_layered1}) and (\ref{diss_perp_layered}). We therefore focus on the first term, which is new and, as mentioned above, may possibly lead to enhanced dissipation. This term yields
\begin{equation}
\label{dis_staggered}
E_s \tau v \rho^2 \int (\pa_{xy} u_x)^2 dx dy \sim E_s \tau v \rho^2 \frac{\sigma^2_\infty L}{E^2_s \rho^3} \int_{x \sim \rho d} \int_{y \sim d} \frac{k\!\left(\frac{x}{\rho~ y}\right)}{\rho y^3} dx dy \ .
\end{equation}
The integral alone yields $d^{-1}$. Therefore, this contribution to the viscous dissipation reads
\begin{equation}
E_s \tau v \rho^2 \int (\pa_{xy} u_x)^2 dx dy \sim E_s \tau v \frac{\sigma^2_\infty L}{E^2_s \rho} \frac{1}{d} \sim G_0 \frac{v \tau}{d}  \ ,
\end{equation}
which is identical to the contribution from the second term. We therefore conclude, somewhat surprisingly, that the viscous dissipation accompanied perpendicular crack propagation in staggered structures are enhanced compared to stratified structures only by a multiplicative factor of order unity, i.e. scaling-wise it remains unchanged
\begin{equation}
G_{vis} \sim G_0 \frac{\tau v}{d}  \ .
\end{equation}
The origin of this unexpected result is that while the dissipation associated with the deformation in the $x$-direction in Eq. (\ref{dis_staggered}) is indeed {\em enhanced} by a large factor $\rho^2$, the displacement $u_x$ in Eq. (\ref{G_perp_stag}) is {\em reduced} in a way that precisely cancels out the $\rho$ dependence of the viscous dissipation.

\section{Anisotropy and fracture criteria}
\label{fracture_criteria}

Our goal here is to demonstrate, through a brief example, that the anisotropy of the structure and the constitutive response may have implications for fracture criteria. The example we focus on is a crack perpendicular to the layers in a stratified structure (i.e. located along the $y$-direction). Using standard fracture mechanics, Eq. (\ref{eom_perp}) immediately implies
\begin{equation}
\sigma_{xx} \sim \frac{\epsilon^{-1/4} \sigma_\infty \sqrt{L}}{(x^2 + \epsilon^{-1} y ^2)^{1/4}} \ .
\end{equation}
The prefactor $\epsilon^{-1/4}$ ensures that $\sigma_{xx}$ approaches $\sigma_\infty$ on a scale $y\!\sim\! L$, as expected.
Focusing on the crack symmetry line, $x\!=\!0$, we obtain
\begin{equation}
\label{stress_perp}
\sigma_{xx}(x\!=\!0,y) \sim \frac{\sigma_\infty \sqrt{L}}{\sqrt{y}} \ .
\end{equation}
This result shows that the near tip stress is {\em unchanged} as compared to isotropic fracture (i.e. the effective stress intensity factor \citep{Broberg1999} remains $K^{eff}\!\sim\!\sigma_\infty \sqrt{L}$). However, the energy release rate for this configuration, already calculated in Eq. (\ref{G_perp}), is significantly {\em enhanced} as compared to isotropic fracture (by a factor $\epsilon^{-1/2}$). This shows that the standard isotropic relation between the stress intensity factor and the energy release rate may not be valid in anisotropic situations. In addition, this comment suggests that some caution should be taken in relating a fracture toughness criterion (i.e. a critical stress intensity factor $K_c$) and a fracture energy criterion (i.e. a critical energy release rate $G_c$) in anisotropic materials.

To further strengthen the point about such anisotropic effects, we briefly consider a related fracture problem in which a long crack (again perpendicular to the layers in a stratified structure) is located on the symmetry line of an infinite strip of width $W$ and loaded by a fixed-grip boundary condition at the lateral edges of the strip. In this case the relevant macroscopic lengthscale in the problem is $W$ and not the crack length (which is assumed to be much larger). This problem was considered in \citet{Okumura2001}. Global energy balance during crack propagation immediately implies that
\begin{equation}
\label{G_strip}
G_0 \sim \frac{\sigma^2_\infty W}{E_h} \ ,
\end{equation}
where $\sigma_\infty$ here is the homogeneous stress far ahead of the crack tip (even though there is no real ``infinity'' here).
\citet{Okumura2001} found that
\begin{equation}
\label{u_strip}
\sigma_{xx}(x\!=\!0,y) \sim \frac{\epsilon^{1/4} \sigma_\infty \sqrt{W}}{\sqrt{y}}\quad\hbox{and}\quad u_x(x\!=\!0,y) \sim \frac{\epsilon^{-1/4} \sigma_\infty \sqrt{W~y}}{E_h} \ ,
\end{equation}
which is indeed consistent with Eq. (\ref{G_strip}) through $G_0\!\sim\!\sigma_{xx} u_x$. Comparing Eqs. (\ref{G_strip}) and (\ref{u_strip}) with Eqs. (\ref{G_perp}) and (\ref{stress_perp}) demonstrates that fracture has different properties in strip geometry and in infinite medium for anisotropic materials. Therefore, again, some caution should be taken in interpreting fracture criteria.

\section{The higher velocities regime}
\label{velocities}

The results presented up to now are restricted to small crack velocities, i.e. they were derived in the framework of a perturbation theory. For example, in the case of stratified and staggered composites, the relevant velocities range is $v\tau/d_\|\!<\!1$ (where $d_\|$ may vary from one problem to another). It is important to note that strictly speaking the perturbative approach is valid for $v\tau/d_\|\!\ll\!1$, which is the small parameter in the perturbative expansion, but as usual the results provide sensible estimates up to $v\tau/d_\|\!\sim\!{\C O}(1)$. Our goal here is to demonstrate, through an example, how one can obtain $G_{vis}(v)$ at finite velocities without solving the equations of motion and evaluating the dissipation integral.

To elucidate the procedure we adopt, we first discuss isotropic viscoelastic fracture. We start with the perturbative result, i.e. Eq. (\ref{dissipation LEFM}), which is presented here again for completeness
\begin{eqnarray}
\label{dissipation LEFM2}
G_{vis} \sim G_0 \frac{v~\tau}{{d_c}} \ .
\end{eqnarray}
As explained above, it is valid for $v\tau/d_c\!<\!1$. Our aim is to extend this result to higher velocities, $v\tau/d_c\!>\!1$. We first note that at higher velocities rate-dependent effects may emerge. That is, regions at different distances from the crack tip, $r\!>\!d_c$, may experience qualitatively different strain rates $\dot\varepsilon$ through the relation
\begin{equation}
\dot\varepsilon = v/r \ .
\end{equation}
In particular, with increasing velocity the region near the tip of a crack may experience strain rates $\dot\varepsilon\!\gg\!\tau^{-1}$ which may result in a modified material response. In fact, we expect disordered (e.g. polymeric) materials to exhibit significant stiffening when deformed at rates much larger than their typical mechanical relaxation rate. Therefore, following \citet{Gennes1996}, we assume that the material under consideration is viscoelastic, characterized by $E_s$ and $\eta_s$, at small strain rates and elastic, characterized by $E_h\!\gg\!E_s$, at large strain rates. For sufficiently large crack velocities a region characterized by $E_h$ develops around the crack tip. Under these conditions $G_{vis}$ has been shown to take the form \citep{Gennes1996}
\begin{equation}
\label{DG}
G_{vis} = G_0 \frac{E_h}{E_s} \ .
\end{equation}
The question is then how the perturbative, small $v$, result of Eq. (\ref{dissipation LEFM2}) is smoothly connected to the large $v$ regime of Eq. (\ref{DG}). This problem was addressed and systematically solved, in a broader viscoelastic context, in \citet{Persson2005}. We follow here the approach of the latter paper. The important observation made in \citep{Persson2005} is that the cutoff scale $d_c$ is in fact a dynamic quantity that grows with $v$, $d_c(v)$, and hence we should interpret all previous appearances of $d_c$ as $d_c(v\!=\!0)$. Furthermore, it was assumed that $d_c(v)$ evolves such that the stress level at that distance from the tip is constant, i.e. that there exists an $v$-independent yield stress or some analogous stress quantity. With this assumption, it was shown in \citet{Persson2005} that
\begin{equation}
\label{PB}
\frac{G_{tot}(v)}{G_0} \sim \frac{d_c(v)}{d_c(0)} \ ,
\end{equation}
where the dissipative contribution emerging from the region $r\!<\!d_c(v)$ (plastic and process zone) was neglected as compared to the viscoelastic contribution emerging from the region $r\!>\!d_c(v)$. While we do not think these assumptions are universally valid, we do believe they provide a possible framework to make progress and gain insight into the fracture dissipation in this class of problems.

With Eqs. (\ref{dissipation LEFM2}), (\ref{DG}) and (\ref{PB}) at hand we can now complete the calculation using ``matched asymptotics'' considerations. We first require that Eq. (\ref{dissipation LEFM2}) is approached in the limit $v\!\to\!0$ and hence replace $d_c$ there by $d_c(v)$. Eliminating $d_c(v)$ between the resulting equation and Eq. (\ref{PB}), using $G_{tot}\!\simeq\!G_{vis}$ and solving for $G_{vis}$, we obtain
\begin{equation}
\label{intermediate}
G_{vis} \sim G_0 \sqrt{\frac{v \tau}{d_c}} \ .
\end{equation}
Since this result should smoothly connect between Eqs. (\ref{dissipation LEFM2}) and (\ref{DG}), it must be valid in the following range of velocities
\begin{equation}
\label{range}
\frac{d_c}{\tau} < v < \left(\frac{E_h}{E_h}\right)^2\frac{d_c}{\tau} \ .
\end{equation}
This provides a description of $G_{vis}(v)$ beyond perturbation theory and over a wide range of crack velocities, in accord with \citet{Persson2005}. We stress that inertia, as in the rest of this paper, is negligible is the present context.

We now demonstrate how these ideas are applied to anisotropic viscoelastic fracture of biological composites. We focus on large cracks that propagate parallel to the layers in a stratified structure and assume that the soft constituent material has a response similar to the material considered above in the isotropic case. The hard material, as elsewhere in this paper, is purely elastic with a modulus $E_h$ at all strain rates. The perturbative result of Eqs. (\ref{dissipation startified1}) and (\ref{d_parallel}) reads
\begin{equation}
\label{d_parallel1}
G_{vis} \sim  G_0 \frac{v \tau}{d_\|} \sim G_0 \frac{v \tau}{\sqrt{d\,\lambda}} \ .
\end{equation}
In this case the small scale $d_\|\!=\!d_x$ is related to the characteristic small scale in the $y$-direction, $d_y$, by the relation $d_x^2\!\sim\!\lambda d_y$. For small velocities, $v\!<\!\sqrt{\lambda\,d}/\tau\!=\!\sqrt{\lambda/d}(d/\tau)$ (i.e. in the perturbative regime), we know that $d_x \!\sim\! \sqrt{\lambda d}$ and $d_y\!\sim\!d$. For larger velocities we assume, as in the isotropic case, that Eq. (\ref{d_parallel1}) still holds with $d_\|\!=\!d_x(v)$ defined such that the stress at this scale remains $v$-independent. This leads to
\begin{equation}
\label{par1}
\frac{G_{vis}(v)}{G_0} \sim \frac{d_y(v)}{d_y(0)} \sim \frac{d_x^2(v)}{d_x^2(0)} \ .
\end{equation}
Using the same procedure as before (i.e. eliminating $d_x(v)$ etc.) we obtain
\begin{equation}
\label{par2}
G_{vis} \sim G_0 \left( \frac{v \tau}{\sqrt{\lambda d}} \right )^{2/3} \ .
\end{equation}
This scaling holds until $d_x \!\sim\! d_y\!\sim\!\lambda$, which corresponds to $v\tau\!\sim\!\lambda ^2/d$, where fracture becomes isotropic. The latter is easily obtained by solving for $d_x(v)$ using Eqs. (\ref{par1}) and (\ref{par2}). Therefore, Eq. (\ref{par2}) is valid in the following range of velocities
\begin{equation}
\sqrt{\frac{\lambda}{d}}\frac{d}{\tau}\!<\!v\!<\!\frac{\lambda^2}{d^2}\frac{d}{\tau} \ .
\end{equation}
For $v\tau\!>\!\lambda ^2/d$ fracture is isotropic and is controlled by the soft constituent material, cf. Eq. (\ref{f_large_stratified}). Therefore the above results for isotropic fracture hold and we should determine the range of validity of each expression using again ``matched asymptotics'' considerations. In particular, for $v\tau\!>\!\lambda ^2/d$ we obtain the isotropic result, cf. Eq. (\ref{intermediate}),
\begin{equation}
\label{par3}
G_{vis} \sim G_0\sqrt{\frac{v\tau}{d}} \ ,
\end{equation}
which is smoothly connected to Eq. (\ref{par2}) at $v\tau\!\sim\!\lambda^2/d$.
This result is valid until $v\tau/d\!\sim\!(E_h/E_s)^2$, where it crosses over to (cf. Eq. (\ref{DG}))
\begin{equation}
\label{DG1}
G_{vis} = G_0 \frac{E_h}{E_s} \ .
\end{equation}
The complete results for the example worked out in detail here are summarized in Table 3. It is easily confirmed, by direct substitution, that the result in each regime smoothly crosses over to the result at the next regime at yet higher velocities. Therefore, we have demonstrated how one can obtain the behavior of $G_{vis}(v)$ for an anisotropic structure for a wide range of velocities, going beyond the theory of perturbation.

\section{Summary and conclusions}
\label{summary}

In this paper we calculated the scaling properties of the energy release rate and the fracture viscous dissipation for various anisotropic composite structures. Our results are summarized in Tables 1-3. The dominant lengthscales for energy release and viscoelastic dissipation can be identified together with the non-trivial scaling behavior emerging from the various anisotropic composite structures.

\begin{table}[here]
\begin{center}
\caption{$G_0$ and $G_{vis}/G_0$ for slow cracks (perturbative regime) parallel to the hard constituent material in the various structures considered in this paper. Recall that $\lambda\!\equiv\!d\sqrt{E_h/E_s}$.}
\begin{tabular}{|c|c|c|c|}
  \hline
  \multicolumn{4}{|c|}{Slow cracks parallel to the hard constituent material}\\ \hline
  & Liquid-crystal-like &  Stratified composites &  Staggered composites \\ \hline
  $G_0$ & $\frac{\sigma_\infty^2 L^2}{E d}$ & $\frac{\sigma_\infty^2 L}{E_s}$ & $\frac{\sigma_\infty^2 L}{E_s}$ \\ \hline
  $G_{vis}/G_0$ & $\frac{v \tau}{d}\frac{L}{d}$ & $\frac{v \tau}{\sqrt{d\,\lambda}}$ & $\frac{v \tau}{\sqrt{d\,\lambda}}$ \\ \hline
\end{tabular}
\end{center}
\end{table}
\begin{table}[here]
\begin{center}
\caption{$G_0$ and $G_{vis}/G_0$ for slow cracks (perturbative regime) perpendicular to the hard constituent material in the various structures considered in this paper. Recall that $\rho$ is the aspect ratio of the hard platelets in the staggered structure.}
\begin{tabular}{|c|c|c|c|}
  \hline
  \multicolumn{4}{|c|}{Slow cracks perpendicular to the hard constituent material}\\
  \hline
  & Liquid-crystal-like &  Stratified composites &  Staggered composites \\ \hline
  $G_0$ & -- & $\frac{\sigma_\infty^2 L}{\sqrt{E_s E_h}}$ & $\frac{\sigma_\infty^2 L}{E_s \rho}$ \\ \hline
  $G_{vis}/G_0$ & -- & $\frac{v \tau}{d}$ & $\frac{v \tau}{d}$ \\
  \hline
\end{tabular}
\end{center}
\end{table}
\begin{table}[here]
\begin{center}
\caption{$G_{vis}/G_0$ for cracks parallel to the hard layers in stratified composites for a wide range of velocities, i.e. in the non-perturbative regime. Recall that $\lambda\!\equiv\!d\sqrt{E_h/E_s}$}
\begin{tabular}{|c|c|c|c|c|}
  \hline
  \multicolumn{5}{|c|}{Cracks parallel to the hard layers in stratified composites}\\ \hline
  Velocity range & $v\!<\!\sqrt{\frac{\lambda}{d}}\frac{d}{\tau}$ &  $\sqrt{\frac{\lambda}{d}}\frac{d}{\tau}\!<\!v\!<\!\frac{\lambda^2}{d^2}\frac{d}{\tau}$ &  $\frac{\lambda^2}{d^2}\frac{\tau}{d}\!<\!v\!<\!\left(\frac{E_h}{E_s}\right)^2\!\frac{d}{\tau}$ & $v\!>\!\left(\frac{E_h}{E_s}\right)^2\frac{d}{\tau}$ \\ \hline
  $G_{vis}/G_0$ & $\frac{v \tau}{\sqrt{\lambda\,d}}$ & $\left(\frac{v \tau}{\sqrt{\lambda d}}\right)^{2/3}$ & $\sqrt{\frac{v\tau}{d}}$ & $\frac{E_h}{E_s}$\\
  \hline
\end{tabular}
\end{center}
\end{table}

We would like to highlight the qualitative difference between the viscoelastic dissipation in liquid-crystal-like composites and the other composites considered in this paper. The dissipation integral for the liquid-crystal-like composites, Eq. (\ref{integral_lc}), is dominated by the upper limits of integration, which is a result of the integrable singularity of the integrand. The consequence is that viscoelastic dissipation in this case comes from a region whose size is controlled by a macroscopic, extrinsic, material-independent, lengthscale. In contrast, the viscoelastic dissipation of other composites discussed here, even when involves different scales as compared to isotropic fracture, is always controlled by intrinsic, material-dependent, lengthscales. This derives from the appearance of a non-integrable singularity in the dissipation integral, which is a generic feature of ordinary fracture. We believe that this qualitative difference emerges from the fact that the anisotropy of the liquid-crystal-like composites is {\em infinite}, i.e. the direction parallel to the layers exhibits no elastic response at all (i.e. it features a liquid response in this direction), while in all other cases the anisotropy may be (very) large, but always {\em finite}.

We believe that our results may have both theoretical and practical merit. From a theoretical point of view, our results show that in the presence of heterogeneous structures and variability in the local mechanical properties, the linear elastic energy release rate and the fracture viscous dissipation may exhibit different properties as compared to their isotropic fracture counterparts. In addition, our work further demonstrates the power and potential of continuum approaches in understanding biological (and other) composite materials. Note, however, that we cannot exclude other important effects emerging from scales that are not resolved in the coarse-grained approach.

From a more practical point of view, our work highlights the possible importance of viscous dissipation as a toughening mechanism in biological composites. This may be at least partially supported by the difference between the fracture toughness of hydrated and dehydrated biological composites. We suspect that this velocity-dependent fracture toughness enhancement may be quantitatively important, in addition to the fracture toughness associated with crack extension \citep{Nalla2004, Launey2010}. However, careful experiments are needed in order to test our suggestion. Such experiments require controlling the applied $G_{tot}$ and accurately measuring the critical conditions at crack initiation $G_0\!=\!G_c$, in order to calculate $G_{vis}/G_0$. Such measurements also involve monitoring the crack velocity $v$ and independently extracting material parameters such as $\tau$ and $E_s/E_h$. At present, we could not find extensive and systematic experimental data on such quantities.

Finally, our results demonstrate the relevance of anisotropy to the interpretation of different fracture criteria in biological composites. This may be useful for interpreting and characterizing the fracture-related material properties of these composites. In this context, it is important to note that we focussed on strong anisotropy (either in elastic or fracture properties), which is indeed observed at small scales \citep{Peterlik2006a}, but usually not on macroscopic ones, cf. \citet{Behiri1989, Nalla2003}. The latter is understood as resulting from additional, larger scale, structures that are common in multi-level hierarchial biological composites \citep{Fratzl2007, Dunlop2010}. This may be the subject of a future investigation.\\

{\bf Acknowledgements}\\
EB acknowledges a useful discussion with Peter Fratzl and the support of the Harold Perlman Family Foundation and the Robert Rees Applied Research Fund. EAB acknowledges support of the Erna and Jacob Michael visiting professorship funds at Weizmann Institute of Science.

\appendix

\section{Exact coarse-grained elastic energy for stratified structures}
\label{appendix}

Here we derive the exact coarse-grained linear elastic response of stratified (layered) structures. The hard material has a Young's modulus $E_h$ and a Poisson ratio $\nu_h$, and a thickness $d_h$. The soft material is characterized by $E_s$ and $\nu_s$, and a thickness $d_s$. The volume fraction of the hard material is given by $\phi\!=\!d_h/(d_h+d_s)$. Let us first derive the effective Hooke's law for this structure. We treat the hard and soft materials as isotropic and linear elastic, and the interfaces between them as perfect. We denote the periodicity direction by $y$ and the perpendicular direction by $x$, and assume plane stress conditions. We therefore obtain \citep{Landau1986}
\begin{eqnarray}
\varepsilon^{(h)}_{yy}&=& \frac{\sigma^{(h)}_{yy}-\nu_h \sigma^{(h)}_{xx}}{E_h},\quad\varepsilon^{(s)}_{yy}= \frac{\sigma^{(s)}_{yy}-\nu_s\sigma^{(s)}_{xx}}{E_s},\quad \varepsilon^{(h)}_{xx}=\frac{\sigma^{(h)}_{xx}-\nu_h \sigma^{(h)}_{yy}}{E_h},\quad \varepsilon^{(s)}_{xx}=\frac{\sigma^{(s)}_{xx}-\nu_s \sigma^{(s)}_{yy}}{E_s} \ ,\\
\varepsilon^{(h)}_{xy}&=&\varepsilon^{(h)}_{yx}= \frac{\sigma^{(h)}_{xy}(1+\nu_h)}{E_h} = \frac{\sigma^{(h)}_{yx}(1+\nu_h)}{E_h},\quad
\varepsilon^{(s)}_{xy}=\varepsilon^{(s)}_{yx}= \frac{\sigma^{(s)}_{xy}(1+\nu_s)}{E_s} = \frac{\sigma^{(s)}_{yx}(1+\nu_s)}{E_s}\ .
\end{eqnarray}
Here we used the definition of the linear elastic strain tensor
\begin{equation}
\varepsilon^{(h,s)}_{ij}= \frac{1}{2}\left(\pa_i u^{(h,s)}_j + \pa_j u^{(h,s)}_i \right)
\end{equation}
and the conservation of angular momentum in each material $\sigma^{(h,s)}_{xy}\!=\! \sigma^{(h,s)}_{yx}$. The superscripts $(s)$ and $(h)$ correspond here to the soft and hard layers, respectively.

The boundary conditions demand that the displacement along the interface is the same in both materials (i.e. a perfect interface) and that the force across the interface is continuous. Since the interface is placed along the $x$-axis (constant $y$), these conditions translate into
\begin{eqnarray}
\pa_x u^{(h)}_x \!=\! \pa_x u^{(s)}_x,\quad\pa_x u^{(h)}_y \!=\! \pa_x u^{(s)}_y, \quad\sigma^{(s)}_{yx} \!=\! \sigma^{(h)}_{yx},\quad\sigma^{(s)}_{yy}\!=\! \sigma^{(h)}_{yy} \ .
\end{eqnarray}
Using angular momentum conservation we also obtain $\sigma^{(s)}_{xy} \!=\! \sigma^{(h)}_{xy}$.
Let us first consider the extensional components. The average $\sigma_{xx}$ is given by $\sigma_{xx}\!=\! (1-\phi) \sigma^{(s)}_{xx}\!+\!\phi \sigma^{(h)}_{xx}$.

The condition $\varepsilon_{xx}\!=\!\varepsilon^{(s)}_{xx}\!=\!\varepsilon^{(h)}_{xx}$ implies
\begin{eqnarray}
\frac{\sigma^{(h)}_{xx}-\nu_h \sigma^{(h)}_{yy}}{E_h} =\frac{\sigma^{(s)}_{xx}-\nu_s \sigma^{(s)}_{yy}}{E_s} \ .
\end{eqnarray}
Solving the last two equations we obtain
\begin{eqnarray}
\sigma^{(s)}_{xx} = \frac{\displaystyle \frac{\sigma_{xx}}{\phi} + \left(\frac{E_h}{E_s}\nu_s-\nu_h \right) \sigma_{yy}}{\displaystyle \frac{E_h}{E_s}+\frac{1-\phi}{\phi}}, \quad
\sigma^{(h)}_{xx} = \frac{\sigma_{xx}-(1-\phi)\sigma^{(s)}_{xx}}{\phi} \ .
\end{eqnarray}
Substituting these expressions in
\begin{eqnarray}
\varepsilon_{yy} = \phi \varepsilon^{(h)}_{yy} + (1-\phi) \varepsilon^{(s)}_{yy} =  \frac{\phi(\sigma_{yy}-\nu_h \sigma^{(h)}_{xx})}{E_h} +  \frac{(1-\phi)(\sigma_{yy}-\nu_s \sigma^{(s)}_{xx})}{E_s}, \quad
\varepsilon_{xx} = \frac{\sigma^{(s)}_{xx}-\nu_s \sigma_{yy}}{E_s}
\end{eqnarray}
results in the extensional part of the effective Hooke's law, that is, a linear relation between $\varepsilon_{xx}, \varepsilon_{yy}$ and $\sigma_{xx}, \sigma_{yy}$. For the shear part we have $\pa_x u_y \!=\! \pa_x u^{(s)}_y \!=\! \pa_x u^{(h)}_y$. Therefore,
\begin{eqnarray}
\frac{E_s \left(\pa_x u_y + \pa_y u^{(s)}_x\right)}{2(1+\nu_s)} = \frac{E_h \left(\pa_x u_y + \pa_y u^{(h)}_x\right)}{2(1+\nu_h)} = \sigma_{xy} \ .
\end{eqnarray}
Expressing $\pa_y u^{(s,h)}_x$ in terms of $\sigma_{xy}$ and using $\pa_y u_x \!=\! (1-\phi)\pa_y u^{(s)}_x \!+\! \phi \pa_y u^{(h)}_x$, we obtain
\begin{eqnarray}
\varepsilon_{xy}=\frac{1}{2}(\pa_x u_y+\pa_y u_x) = \left[\frac{(1-\phi)(1+\nu_s)}{E_s} + \frac{\phi(1+\nu_h)}{E_h} \right] \sigma_{xy} \ .
\end{eqnarray}

Using these expressions, the elastic energy density $\varepsilon_{ij}\sigma_{ij}/2$ (excluding the bending energy) can be immediately calculated. Assuming $E_h \gg E_s$ and $\phi \simeq 1/2$, adding the bending energy and omitting prefactors of order unity, we obtain Eq. (\ref{elastic_layered}) in the text.

\bibliographystyle{elsart-harv}
\bibliography{Biomaterials}

\begin{thebibliography}{37}
\expandafter\ifx\csname natexlab\endcsname\relax\def\natexlab#1{#1}\fi
\expandafter\ifx\csname url\endcsname\relax
  \def\url#1{\texttt{#1}}\fi
\expandafter\ifx\csname urlprefix\endcsname\relax\def\urlprefix{URL }\fi

\bibitem[{Antonietti and Fratzl(2010)}]{Antonietti2010}
Antonietti, M., Fratzl, P., Jan. 2010. {Biomimetic Principles in Polymer and
  Material Science}. Macromolecular Chemistry and Physics 211~(2), 166--170.

\bibitem[{Arzt et~al.(2003)Arzt, Fratzl, Gao, Ji, and Ja}]{Arzt2003}
Arzt, E., Fratzl, P., Gao, H., Ji, B., Ja, I.~L., 2003. {Materials become
  insensitive to flaws at nanoscale: Lessons from nature}. Proceedings of the
  National Academy of Sciences of the United States of America 100~(10),
  5597--5600.

\bibitem[{Behiri and Bonfield(1989)}]{Behiri1989}
Behiri, J.~C., Bonfield, W., 1989. {Orientation dependence of the fracture
  mechanics of cortical bone.} Journal of biomechanics 22~(8-9), 863--72.

\bibitem[{Broberg(1999)}]{Broberg1999}
Broberg, K.~B., 1999. Cracks and Fracture. Academic Press, San Diego.

\bibitem[{de~Gennes(1990)}]{Gennes1990}
de~Gennes, P.~G., 1990. {Lenticular Fracture of a Smectic Liquid Crystal}.
  Europhysics Letters (EPL) 13~(8), 709--714.

\bibitem[{de~Gennes(1996)}]{Gennes1996}
de~Gennes, P.~G., 1996. {Soft Adhesives}. Langmuir 12~(19), 4497--4500.

\bibitem[{de~Gennes(2000)}]{Gennes2000}
de~Gennes, P.~G., 2000. {On the toughness of biocomposites}. Comptes Rendus de
  l'Acad\'{e}mie des Sciences - Series IV - Physics 1~(2), 257--261.

\bibitem[{Dunlop and Fratzl(2010)}]{Dunlop2010}
Dunlop, J.~W., Fratzl, P., 2010. {Biological Composites}. Annual Review of
  Materials Research 40~(1), 1--24.

\bibitem[{Fratzl(2007)}]{Fratzl2007}
Fratzl, P., 2007. {Biomimetic materials research: what can we really learn from
  nature's structural materials?} Journal of the Royal Society Interface
  4~(15), 637--42.

\bibitem[{Fratzl and Weinkamer(2007)}]{Fratzl2007a}
Fratzl, P., Weinkamer, R., 2007. {Nature's hierarchical materials}. Progress in
  Materials Science 52, 1263--1334.

\bibitem[{Gao(2006)}]{Gao2006}
Gao, H., 2006. {Application of Fracture Mechanics Concepts to Hierarchical
  Biomechanics of Bone and Bone-like Materials}. International Journal of
  Fracture 138~(1-4), 101--137.

\bibitem[{Hazenberg et~al.(2006)Hazenberg, Taylor, and {Clive
  Lee}}]{Hazenberg2006}
Hazenberg, J.~G., Taylor, D., {Clive Lee}, T., 2006. {Mechanisms of short crack
  growth at constant stress in bone.} Biomaterials 27~(9), 2114--22.

\bibitem[{Iyo et~al.(2004)Iyo, Maki, Sasaki, and Nakata}]{Iyo2004}
Iyo, T., Maki, Y., Sasaki, N., Nakata, M., 2004. {Anisotropic viscoelastic
  properties of cortical bone.} Journal of biomechanics 37~(9), 1433--7.

\bibitem[{Ji and Gao(2004)}]{Ji2004}
Ji, B., Gao, H., 2004. {Mechanical properties of nanostructure of biological
  materials}. Journal of the Mechanics and Physics of Solids 52, 1963 -- 1990.

\bibitem[{Ji and Gao(2010)}]{Ji2010}
Ji, B., Gao, H., 2010. {Mechanical Principles of Biological Nanocomposites}.
  Annual Review of Materials Research 40, 77--100.

\bibitem[{Kato(2002)}]{Kato2002}
Kato, T., 2002. {Self-Assembly of Phase-Segregated Liquid Crystal Structures}.
  Science 295, 2414--2418.

\bibitem[{Kinney and Ritchie(2005)}]{Kinney2005}
Kinney, J.~H., Ritchie, R.~O., 2005. {Fracture in human cortical bone: local
  fracture criteria and toughening mechanisms}. Journal of Biomechanics 38,
  1517--1525.

\bibitem[{Koester et~al.(2008)Koester, Ager, and Ritchie}]{Koester2008}
Koester, K.~J., Ager, J.~W., Ritchie, R.~O., 2008. {Aging and fracture of human
  cortical bone and tooth dentin}. Jom 60~(6), 33--38.

\bibitem[{Kruzic(2003)}]{Kruzic2003}
Kruzic, J., 2003. {Crack blunting, crack bridging and resistance-curve fracture
  mechanics in dentin: effect of hydration}. Biomaterials 24~(28), 5209--5221.

\bibitem[{Landau and Lifshitz(1986)}]{Landau1986}
Landau, L.~D., Lifshitz, E.~M., 1986. Theory of Elasticity, 3rd Edition.
  Pergamon Press, London.

\bibitem[{Launey and Ritchie(2009)}]{Launey2009}
Launey, B. M.~E., Ritchie, R.~O., 2009. {On the Fracture Toughness of Advanced
  Materials} 94720, 2103--2110.

\bibitem[{Launey et~al.(2010)Launey, Buehler, and Ritchie}]{Launey2010}
Launey, M.~E., Buehler, M.~J., Ritchie, R.~O., 2010. {On the Mechanistic
  Origins of Toughness in Bone}. Annual Review of Materials
  Research~(February), 1--29.

\bibitem[{Munch et~al.(2008)Munch, Launey, Alsem, Saiz, Tomsia, and
  Ritchie}]{Munch2008}
Munch, E., Launey, M., Alsem, D., Saiz, E., Tomsia, A., Ritchie, R., 2008.
  Tough bio-inspired hybrid materials. Science 322, 1516--1520.

\bibitem[{Nalla et~al.(2003)Nalla, Kinney, and Ritchie}]{Nalla2003}
Nalla, R., Kinney, J., Ritchie, R., 2003. {Effect of orientation on the in
  vitro fracture toughness of dentin: the role of toughening mechanisms}.
  Biomaterials 24~(22), 3955--3968.

\bibitem[{Nalla et~al.(2005)Nalla, Kruzic, Kinney, and Ritchie}]{Nalla2005}
Nalla, R.~K., Kruzic, J.~J., Kinney, J.~H., Ritchie, R.~O., 2005. {Mechanistic
  aspects of fracture and R-curve behavior in human cortical bone.}
  Biomaterials 26~(2), 217--31.

\bibitem[{Nalla et~al.(2004)Nalla, Kruzic, and Ritchie}]{Nalla2004}
Nalla, R.~K., Kruzic, J.~J., Ritchie, R.~O., 2004. {On the origin of the
  toughness of mineralized tissue: microcracking or crack bridging?} Bone 34,
  790 -- 798.

\bibitem[{Okumura(2002{\natexlab{a}})}]{Okumura2002a}
Okumura, K., 2002{\natexlab{a}}. {Physical picture for fractures in stratified
  materials: viscoelastic effects in large cracks}, arXiv:cond--mat/0212532v1.

\bibitem[{Okumura(2002{\natexlab{b}})}]{Okumura2002}
Okumura, K., 2002{\natexlab{b}}. {Why is nacre strong? II. Remaining mechanical
  weakness for cracks propagating along the sheets}. The European Physical
  Journal E 310, 303--310.

\bibitem[{Okumura(2003)}]{Okumura2003}
Okumura, K., 2003. {Enhanced energy of parallel fractures in nacre-like
  composite materials}. Europhysics Letters (EPL) 63~(5), 701--707.

\bibitem[{Okumura(2005)}]{Okumura2005}
Okumura, K., 2005. {Fracture strength of biomimetic composites: scaling views
  on nacre}. Journal of Physics: Condensed Matter 17~(31), S2879--S2884.

\bibitem[{Okumura and de~Gennes(2001)}]{Okumura2001}
Okumura, K., de~Gennes, P.-G., 2001. {Why is nacre strong? Elastic theory and
  fracture mechanics for biocomposites with stratified structures}. The
  European Physical Journal E 4~(1), 121--127.

\bibitem[{Persson and Brener(2005)}]{Persson2005}
Persson, B., Brener, E., 2005. {Crack propagation in viscoelastic solids}.
  Physical Review E 71~(3), 1--8.

\bibitem[{Peterlik et~al.(2006)Peterlik, Roschger, Klaushofer, and
  Fratzl}]{Peterlik2006a}
Peterlik, H., Roschger, P., Klaushofer, K., Fratzl, P., 2006. {From brittle to
  ductile fracture of bone.} Nature materials 5~(1), 52--5.

\bibitem[{Puxkandl et~al.(2002)Puxkandl, Zizak, Paris, Keckes, Tesch,
  Bernstorff, Purslow, and Fratzl}]{Puxkandl2002}
Puxkandl, R., Zizak, I., Paris, O., Keckes, J., Tesch, W., Bernstorff, S.,
  Purslow, P., Fratzl, P., 2002. {Viscoelastic properties of collagen:
  synchrotron radiation investigations and structural model.} Philosophical
  transactions of the Royal Society of London. Series B, Biological sciences
  357~(1418), 191--7.

\bibitem[{Ritchie(2010)}]{Ritchie2010}
Ritchie, R.~O., 2010. {How does human bone resist fracture?} Annals of the New
  York Academy of Sciences 1192, 72--80.

\bibitem[{Ritchie et~al.(2009)Ritchie, Buehler, and Hansma}]{Ritchie2009}
Ritchie, R.~O., Buehler, M.~J., Hansma, P., 2009. {Plasticity and toughness in
  bone}. Physics Today, 41--47.

\bibitem[{Sasaki et~al.(1993)Sasaki, Nakayamaa, Yoshikawaa, and
  Enyo}]{Sasaki1993}
Sasaki, N., Nakayamaa, Y., Yoshikawaa, M., Enyo, A., 1993. {Stress Relaxation
  Function of Bone and Bone Collagen} 26~(12), 1369--1376.

\end{thebibliography}

\end{document}